# Estimating body mass in leatherback turtles *Dermochelys coriacea*

Jean-Yves Georges[1*] & Sabrina Fossette[1,2]

[1] Centre National de la Recherche Scientifique, Institut Pluridisciplinaire Hubert Curien, Département d'Ecologie, Physiologie et Ethologie, 23 rue Becquerel, 67087 Strasbourg, France

[2] Université Louis Pasteur, 4 rue Blaise Pascal, 67070 Strasbourg, France

* Corresponding author: Jean-Yves Georges

Centre National de la Recherche Scientifique

Institut Pluridisciplinaire Hubert Curien

Département d'Ecologie, Physiologie et Ethologie

23 rue Becquerel

67087 Strasbourg, France

Phone: +33 388 106 947

Fax: +33 388 106 906

Email: jean-yves.georges@c-strasbourg.fr

**Abstract**

Body mass is a major life history trait and provides a scale for all living processes of organisms. Unfortunately body mass cannot be easily measured for many species, because of the logistical difficulties involved in actually catching and weighing them. This is particularly true for sea turtles which are large vertebrates that spend most of their life at sea. Here we developed a general linear model to predict body mass from 17 morphometric measurements obtained from 49 leatherback turtles nesting in Awala Yalimapo beach, French Guiana, South America. A stepwise backward analysis removing independent parameters with $P > 0.001$ indicated that body mass can be estimated with 93% accuracy by Body Mass (kg) = - 709.146 + 3.391MedianBodyCirc (cm) + 2.664SCCL (cm), where SCCL is the standard curvilinear carapace length and MedianBodyCirc is body circumference at half of SCCL. A sensibility test showed that this parsimonious model is robust, as estimated body mass may change by 0.7-1.3% for 1-2 cm changes in SCCL, by 0.9-1.7% for 1-2 cm changes in circumference. Leatherback turtles from French Guiana were larger and heavier than in all other nesting sites studied so far, suggesting either that gravid leatherbacks feed during the nesting season in French Guiana, and/or that this species may exhibits site-specific growth strategies. Further studies are required to test these hypotheses, including implementation of similar simple models for other nesting populations, in order to better understand the life history of this endangered species.

**Introduction**

Body mass is a major life history trait (Stearns 1992) and provides a scale for all living processes of organisms (Schmidt-Nielsen 1984) including their interactions with their habitat. For example, body mass contributes to the capacity with which organisms can buffer changes in food availability. A larger body mass enables species and/or individuals to better buffer a substantial decrease in food abundance compared with smaller species and/or individuals. This has important consequences for the way organisms reproduce. In animals, larger species have been reported to store body reserves prior to reproduction and rely on them during reproduction (capital breeders, Jonnson 1997). In contrast, smaller species adjust their food intake during the breeding season according to requirements associated with reproduction (income breeders, Jonnson 1997). Body mass thus provides valuable information concerning the potential survival and the reproductive success of a population (Clutton-Brock 1988, Gaillard et al. 2000).

Unfortunately, body mass may be difficult to be measured for many species, particularly for large marine vertebrates such as sea turtles and marine mammals, which spend most of, if not their entire, life at sea and whose adult body mass frequently exceeds several hundred kilograms. Although mean body mass is often estimated from life tables and growth curves (e.g. Trites & Pauly 1998), such life history data do not consistently exist for sea turtles. A few, rare studies investigated the relationship between body mass and morphometrics in sea turtles (e.g. Boulon et al. 1996, Leslie et al. 1996, Hays et al. 2002). These studies focused on body length, probably because it is the most common and one of the easiest measurement to be taken in the field. However, the weak relationship so far reported between body mass and body length indicates that body length is not a good predictor of body mass in sea turtles, particularly in leatherback turtles (Boulon et al. 1996, Leslie et al. 1996). Hence, accurate estimates of key life history traits, such as body mass are crucial for improving our knowledge of the basic biology of this critically endangered species.

Here we propose a method for accurately estimating body mass in gravid leatherback turtles taking simple measurements that can be easily achieved in the field at low cost. We developed a sensibility test to take into account that field work in sea turtles is performed at night, potentially resulting in errors of these field measurements. Finally we compare our results with previous studies and suggest new areas of investigation in sea turtle research.

**Material and Methods**

This study was conducted during the entire 2005 nesting season at Awala Yalimapo beach (5.7°N – 53.9°W), French Guiana, South America. Awala Yalimapo beach is situated on the French side of the Maroni River, separating French Guiana and Surinam, where more than 50% of the world population of leatherback turtles nest (Girondot & Fretey 1996, Spotila et al. 1996). Monitoring programs at this beach have been in place since the late 1970s (Girondot & Fretey 1996, Rivalan et al. 2005a). Nesting individuals have been extensively tagged using external metal tags (Monel tags, National Band and Tags, Newport, Kentucky, USA) and, more recently, with internal Passive Integrated Transponders tags (Trovan Euroid, Courbevoie, France), allowing individual identification. All individuals we considered in the present study were already tagged, so that their identity and their history were known.

Between April 4 and July 21, 2005, a 4 km stretch of the beach where most of nesting events occur (Girondot & Fretey 1996), were continuously patrolled every night from 18:00 to 07:00 hours. All tagged turtles encountered during these patrols were individually identified to ensure that individuals used in the present analyses were considered only once. Additionally, for individuals captured more than once during the study season, we only considered measurements collected during the first capture except for body size (see below). This was necessary to avoid pseudo-replication for statistical analysis. Turtles were measured during ovi-position and weighed afterwards, when returning to the sea. This ensured that disturbance was minimal to the turtles, as suggested by the observation that no turtle ended ovi-position prematurely and all of them came back at least once for a subsequent laying.

In total, 182 individual turtles were opportunistically sampled during our patrolling. Among these 182 individuals, 72 turtles were captured more than once throughout the study period. This permitted to estimate the reliability of our measurements of conservative biometric parameters, such as body length, in order to develop a test of sensibility of our models. During the first month, 49 individuals were captured and measurements of 17 biometrical parameters (see **Figure 1** for details) were obtained using a flexible measuring tape (± 0.5 cm), except head and body heights were obtained using a custom-made calliper rule (± 1cm) when the turtle was lifted for weighing. This permitted to develop in the field a first model identifying the most relevant biometric measurements to be performed during the rest of the study while reducing substantially the time spent per animal. All turtles considered in this study had an intact caudal peduncle to avoid potential biases in body size measurements (Godfrey & Drif 2002). Standard curvilinear carapace length (hereafter referred to SCCL) was measured on the midline of the shell, from the nuchal notch of the carapace to the end of the caudal peduncle (**Figure 1**). For individual turtles captured more than once during the season, we averaged multiple measurements of SCCL to obtain the most accurate value for body length for the analyses, assuming that no detectable linear growth occurred during the nesting season (Broderick et al. 2003, Price et al. 2004, Georges et al. unpublished data). Body circumference measurements were obtained during ovi-position by passing the flexible measuring tape through narrow tunnels below the turtle that were burrowed by hand. Thanks to consistency in the shape of the carapace among individuals, circumferences were measured at the maximum body width (Anterior circumference), at half of the SCCL by marking the midway point when measuring SCCL as a guideline (Median Body Circumference) and at the frontal insertion point of the rear flippers (Posterior Body Circumference) for consistency among individuals. As each turtle was returning to the sea, a custom made harness with four 10-cm wide nylon straps was placed on the sand in front of the turtle, which was secured when the turtle was lying on it. A 4.5 m tall carbon-fibre tripod was immediately placed above the securely harnessed turtle. The turtle was then lifted using a hoist on which a electronic spring scale (600 ± 0.1 Kg, Kern Ltd, Germany) was fixed for weighing. The spring scale was calibrated against a known weight before at the start and at the end of the experiment, and did not show any deviance through time. Additionally, a second measurement of body circumferences was made when the turtle was lifted and did not differ significantly from the measurement made through the tunnels during ovi-position.

During the rest of the nesting season additional turtles were weighed and measured to increase sample size for further analyses. Morphometric measurements (including tunnel digging) and weighing (including harness attachment, turtle lifting and release) did not exceed 10 minutes in total.

Models developed in this study were statistically compared using information statistics rather than simply comparing models for lack of fit according to the Akaike's information criterion (AIC, Burnham & Anderson 1998).

**Results**

Estimating body mass

Between April 4 and 29, 2005, a total of 49 females were extensively measured when laying eggs and weighed after nesting. Body mass after ovi-position (hereafter referred to body mass) ranged from 275.6 to 562.7 kg (mean ± SD: 408.1 ± 63.4 kg, n = 49 turtles). A stepwise backward analysis was performed on a general linear model (GLM) incorporating 17 morphometric variables (**Figure 1**) to identify the most parsimonious model explaining body mass. A first GLM was performed followed by a stepwise backward analysis indicated that the best and most simple model estimating body mass was

Body Mass (kg) = - 699.352 + 3.469MedianBodyCirc (cm) + 2.481SCCL (cm), $F_{2,46}$ = 319.16, $r^2$ = 0.933, P < 0.001, AIC = 421, [**model 1**] where SCCL is the standard curvilinear carapace length (159.1 ± 7.0 cm, range = [146 - 179 cm], n = 49) and MedianBodyCirc is the body circumference at half of SCCL (205.5 ± 14.6 cm, range = [173 – 237 cm], n = 49 **Figure 1** and **Table 1**).

MedianBodyCirc and SCCL accounted for 75%, and 25% of the variance of **model 1**, respectively. Comparisons of Akaike weights indicate that **model 1** was significantly better than the simple linear regression between body mass (y, in kg) and SCCL (x, in cm: y = 6.217x - 580.665, $r^2$ = 0.467, P < 0.001, AIC = 521, P > 0.99), or between body mass and $SCCL^3$ (y = 8.054E$^{-5}$x + 82.155, $r^2$ = 0.458, P < 0.001, AIC 522, P > 0.99) commonly reported in previous studies (e.g. Boulon et al. 1996, Leslie et al. 1996, James et al. 2005), but also between body mass and MedianBodyCirc (y = 4.08x – 430.402, $r^2$ = 0.878, P < 0.001, AIC = 449, P > 0.99).

In order to increase sample size and analysis power, we then assumed that body mass obtained in 182 turtles (which included the 49 individuals used for **model 1**) measured throughout the nesting season was related to MedianBodyCirc and SCCL. Body mass measured throughout the nesting season ranged from 275.6 to 567.3 kg (mean ± SD: 389.7 ± 61.9 kg, n = 182) and was significantly related to MedianBodyCirc and SCCL according to the following equation:

Body Mass (kg) = - 709.146 + 3.391MedianBodyCirc (cm) + 2.664SCCL (cm), $F_{2,179}$ = 1054.476, $r^2$ = 0.922, P < 0.001, AIC = 1563, [**model 2**],

with SCCL (159.3 ± 6.8 cm, range = [142 – 179 cm], n = 182) and MedianBodyCirc (198.9 ± 14.0 cm, range = [170 – 237 cm], n = 182) accounting for 28% and 72% of the variance of the model, respectively.

As for **model 1**, **model 2** calculated throughout the nesting season was significantly better than the simple linear regression between body mass and SCCL (y = 6.517x - 648.268, $r^2$ = 0.514, n = 182, P < 0.001, AIC = 1998, P > 0.99), or between body mass and $SCCL^3$ (y = 9E$^{-5}$x + 39.458, $r^2$ = 0.52, n = 182, P < 0.001, AIC = 1891, P > 0.99), or between body mass and MedianBodyCirc (y = 4.107x – 427.255, $r^2$ = 0.862, P < 0.001, AIC = 1664, P > 0.99).

Body mass estimated from **model 1** ($BMest_1$) was strongly correlated with body mass estimated from **model 2** ($BMest_2$): $BMest_2$ = 0.997$BMest_1$ + 4.762 (**Figure 3**, $F_{1,180}$ = 521390, $r^2$ = 0.9997, P < 0.001). The slope of the regression did not differ from 1 (t-test = 0.004, P > 0.99) but the intercept with the origin differed significantly from 0 (t = 345, P < 0.001), indicating a slight (1%) difference when considering individuals early in the season compared to individuals sampled throughout the nesting season.

Sensibility test

Among the 182 individual turtles considered in this study, 72 of them have been captured more than once. This enabled to estimate the reliability of our measurements, considering that misreading can occur particularly when field work is performed at night, as is the case when working with nesting sea turtles. Measurements of the SCCL made on individual turtles during at least two consecutive nesting events differed by 1.5 ± 1.2 cm (n = 209 measurements performed on 72 turtles), ranging from zero up to 5 cm in some cases. We thus implemented a sensibility test to assess how changes in MedianBodyCirc and/or in SCCL would affect body mass estimates based on **model 2**, considering an hypothetical turtle of 160 cm SCCL and 200 cm MedianBodyCirc, and 395.3 kg (according to **model 2**). Not surprisingly, body mass estimates from **model 2** were more affected by changes in MedianBodyCirc than by changes in SCCL (**Figure 4**). Changes of 1, 2, and 5 cm in MedianBodyCirc around 200 cm resulted in relative variations in body mass of 0.9, 1.7, and 4.3%, respectively (corresponding to 3.4, 6.8, and 16.9 kg, respectively). Similar changes in SCCL around 160 cm would result in relative variations in estimated body mass of 0.7, 1.3, and 3.3%, respectively (i.e. 2.7, 5.7, 13.3 kg, respectively). Relative variations in estimated body mass associated with concurrent changes in MedianBodyCirc and SCCL were additive, i.e. concurrent changes of 1, 2, and 5 cm around 200 cm of MedianBodyCirc and 160 cm of SCCL would result in relative variations in body mass of 1.5, 3.1, and 7.7%, respectively.

**Discussion**

Leatherback turtles rank among the largest living reptiles (Eckert & Luginbuhl 1988), making them logistically difficult to weigh. This problem is compounded by the fact that only females are accessible on land and then only for a period of about 24 hours in total every 2 to 3 years while nesting (Girondot & Fretey 1996, Rivalan et al. 2005b). Body mass is however a key life history trait and has been reported as the most appropriate parameter for assessing somatic growth, superior to the commonly used measurement of carapace length (Bjorndal & Bolten 1988, Chaloupka & Musick 1997, Price et al. 2004). Body mass is also required for estimating food consumption of a population (e.g. Trites et al. 1997) and to assess the trophic rule of a species on its food web. Body mass is thus required for assessing energy fluxes through trophic levels, and for ultimately better understand ecosystem functioning.

The aim of this study was to provide a method, easily applicable in the field by both specialists and non specialists, to accurately estimate body mass in leatherback turtles. We found that in French Guiana, body mass of gravid leatherback turtles, individually captured throughout the nesting season, can robustly be estimated with 93% accuracy from standard curvilinear carapace length (SCCL) and body circumference at half of SCCL (MedianBodyCirc). This model may apply for male individuals as well, as James et al. (2005) found no sex-specific relationship between body mass and SCCL between 7 females and 6 males feeding off Canada. Therefore additional data are needed to establish whether this relationship also holds true for males in French Guiana. Both SCCL and MedianBodyCirc can be taken with relative ease using flexible measuring tape. Body mass was mainly predicted by body circumference, a parameter simple to measure by digging a narrow tunnel below the turtle's body. Although, body circumference can only be measured when the turtle is almost immobile during nest digging and ovi-position, these activities occur over a span of approximately 50 minutes (Billes 2000), allowing more than sufficient time to dig a tunnel. SCCL is the most commonly measured parameter in all sea turtle monitoring programs throughout the world because of its practicability, yet it contributed to less than 30% of our model, and explained only about 50% of the variance of body mass when considered alone. Similar low r² values have previously been reported for body mass – SCCL regressions of gravid leatherback turtles nesting in St Croix Island, Caribbean Sea, and along the Atlantic coast of Costa Rica ($r^2 = 0.551$, and $r^2 = 0.704$, respectively; Boulon et al. 1996, Leslie et al. 1996), indicating that accurate estimates of body mass in leatherback turtles require more than a single parameter. Our study proposes a simple, yet robust and powerful way (**Figure 3**) to improve such estimates of body mass. The fact that slopes and intercepts of body mass − SCCL regressions are slightly different between nesting populations (**Figure 5**, Boulon et al. 1996, Leslie et al. 1996, Southwood et al. 1999, this study) emphasises the importance of similar studies for other nesting sites to test if body mass in other populations can also be estimated from relative simple models.

Differences in the relationship of body mass and SCCL between studies may result from different environmental trophic conditions and/or from different growth strategies. Atlantic leatherback turtles were 18-25% heavier (for a given length) when feeding off Canada than when nesting (**Figure 5**). This is consistent with the fact that gravid sea turtles are commonly considered as not feeding substantially during the nesting season (Miller 1997). Heavier individuals in cold waters may also result from water retention related to osmotic regulation and to the optimisation of body shape in this extended migrant. Interestingly, leatherback turtles nesting in French Guiana in 2005 were about 70 to 120 kg heavier than those from other nesting sites (**Figure 5**, Boulon et al. 1996, Leslie et al. 1996, Southwood et al. 1999) and than individuals weighed in French Guiana in 1987 (Girondot & Fretey 1996). Our study indicates that for an average leatherback turtle of 160 cm SCCL and 200 cm body circumference, changes of 5 cm in SCCL (all remaining similar otherwise) results in a change in body mass of about 13 kg (**Figure 4**). If this tendency holds for other populations too, the large differences in body mass observed between populations may not only be attributed to the slight inter-population differences in SCCL, but are more likely to result from different growth patterns among populations. However, growth patterns are poorly documented in leatherback turtles (but see Zug & Parham 1996, Price et al. 2004) and should be further investigated so that we can test the hypothesis of population-specific growth patterns, but also to test potential sex-specific growth patterns.

Interestingly, the slope of the body mass – SCCL regression we found for leatherbacks nesting in French Guiana was more similar to the slope for individuals feeding off Canada (Body mass [kg] = 6.8CCL [cm] - 600, estimated from James et al. 2005) than to the slopes previously reported for other nesting Atlantic populations (St Croix Island, Caribbean Sea: Body mass [kg] = 5.208SCCL [cm] + 468.84, Boulon et al. 1996; Atlantic coast of Costa Rica: Body mass [kg] = 5.032SCCL [cm] - 417.7, Leslie et al. 1996). This suggests that leatherback turtles nesting in French Guiana feed during the nesting season but to a lesser extend than turtles observed in their feeding grounds off Canada by James et al. (2005) and in other sites, turtles may not feed during nesting or to a lesser extent than in French Guiana. Although this analysis may provide indirect evidence that leatherback turtles may feed during the nesting season in French Guiana (Georges et al. submitted, Fossette et al. submitted) further investigations using direct measurements of body mass changes during the nesting season are required to provide confirmation. In addition, leatherbacks feeding off Canada were on average about 7-8 cm smaller than individuals nesting in French Guiana (estimated from James et al. 2005, this study), yet within the range of body lengths observed in the nesting population of French Guiana. This could also suggest that relatively small leatherbacks observed feeding off Canada might be individuals that originated from, and/or nested in French Guiana. This is supported by recent satellite tracking data (Ferraroli et al. 2004, Georges et al. unpublished data) but further investigations, including genetic analyses, are required to confirm the actual origin of sea turtles opportunistically or accidentally captured on their feeding grounds.

Finally, the relatively large size of gravid leatherback turtles in French Guiana compared to other sites may permit them to better buffer potential changes in food availability during their trans-oceanic migration (Ferraroli et al. 2004). This may contribute to the relatively healthy population in French Guiana and possibly on the Guiana shield (Guiana, Suriname, French Guiana), compared to the declining population in the Pacific Ocean (Spotila et al. 2000), where the smallest specimens have been reported (**Figure 5**). In addition, leatherbacks measured in our study were about 5 cm longer than individuals measured on the same site 18 years earlier (154.6 ± 9.0 cm, n = 15, Girondot & Fretey 1996). Such size difference is consistent with the growth rate estimated to about 0.2 cm/year in adult gravid leatherbacks (Price et al. 2004). Although size is not a good predictor of age in sea turtles, the relatively large size of individuals observed in French Guiana during this study may thus be explained by a relatively high proportion of old individuals within this population. This would imply a different population structure which may also cause different dynamics within this population, compared with other sites.

Managing sea turtle populations requires estimates of vital parameters (growth, mortality, fertility, etc.) that are either hard or impossible to obtain by direct sampling. Greater effort should therefore be directed toward deriving empirical models from easy-to-measure parameters to obtain estimates of parameters which are hard-to measure (such as mean body mass of an individual in an age-structured population). Improving estimation of other parameters (related to physiology, morphology or population dynamics) in leatherback turtles should lead to a better understanding of their life history, trophic ecology and population dynamics, which is urgently needed for developing more appropriate conservation strategies to preserve this critically endangered species.

**Acknowledgements**

We thank G Alberti, M-H Baur, C Blanc, M Bouteille, N Lapompe-Paironne and I Van der Auwera for their assistance in the field, M Girondot for statistical advices and M Enstipp for correcting the English. We are grateful to all Awala-Yalimapo inhabitants for their hospitality and all people involved in the sea turtle monitoring programmes through Kulalasi NGO and the Réserve Naturelle de l'Amana for their logistical contribution in the field. Funding was provided by grants to JYG from Programme Amazonie du CNRS. SF was supported by a studentship from the French Ministry of Research. This study was carried out under CNRS-CEPE institutional license (B67-482 18) and JYG individual license (67-220 and 04-199) and adhered to the legal requirements of the country in which the work was carried out, and all institutional guidelines.

Table 1. Analysis (general linear model) of body mass in gravid leatherback turtles in relation to 17 morphometric variables (see **Figure 1**, complete model : $F_{17,32} = 87.869$, $r^2 = 0.98$, $P < 0.001$). Analysis was run for 49 individuals sampled once when nesting in French Guiana between April 4 and 29, 2005.

| Independent variable | t-value | P-value | b* |
|---|---|---|---|
| **Constant** | **-13.557** | **0.000** | **-699.4** |
| Head length | -1.311 | 0.200 | - |
| Head height | -1.268 | 0.214 | - |
| Head circumference | -0.786 | 0.438 | - |
| Neck length | 0.181 | 0.858 | - |
| Neck circumference | 2.424 | 0.021 | - |
| Anterior Carapace width | 2.983 | 0.006 | - |
| Median Carapace width | -1.662 | 0.107 | - |
| Posterior Carapace width | -0.850 | 0.402 | - |
| **[Curvilinear Carapace Length]** | **4.430** | **0.000** | **2.481** |
| Tail length | 1.270 | 0.213 | - |
| Peduncle length | -2.718 | 0.011 | - |
| Anterior body height | 2.105 | 0.044 | - |
| Median body heigth | -1.236 | 0.226 | - |
| Posterior body height | 0.466 | 0.644 | - |
| Anterior body circumference | 1.787 | 0.084 | - |
| **Median body circumference** | **6.005** | **0.000** | **3.469** |
| Posterior body circumference | -0.229 | 0.820 | - |

**Table 1 continued**

* slope value (only given for significant effects after stepwise backward analysis, removing independent variables for P < 0.0001) with Body mass in kg, and SCCL and MedianBodyCirc in cm (final model : $F_{2,46}$ = 319.160, r² = 0.933, P < 0.001)

**Figure 1.** Schematic representation of morphometric measurements collected on gravid leatherback turtles (left side) nesting in French Guiana. The carapace is presented in solid lines, in contrast to the soft tissues that are shown in dashed lines, except for the tail (filled in black) which was considered in the analysis. Numbers in classic and *in italic* types correspond to straight and *curvilinear* measurements, respectively. **1** head length, from the distal point of the nose to the tip of occipital bone; **2-3** head height and head circumference at the level of the occipital bone; **4** neck length from the occipital bone to the nape notch of the carapace; **5** neck circumference, measured at the maximum circumference during apnea, **6** anterior carapace width at the level of maximum of maximum width back of the fore-flippers, **7** median carapace width at half of 9 (SCCL), **8** posterior carapace width at the frontal insertion point of the rear-flippers, *9* standard curvilinear carapace length (SCCL) measured on the midline of the shell, from the nape notch of the carapace to the distal point of the peduncle, **10** tail length measured from below of the tail when stretched during ovi-position, **11** peduncle length measured from below of the peduncle from the posterior end of the soft body to the distal point of the peduncle, **12-13-14** anterior-median-posterior body height, respectively, measured at the same levels as *6-7-8* when the turtle was lifted for weighing, *15-16-17* anterior-median-posterior body circumference, respectively, measured at the same levels as *6-7-8*.

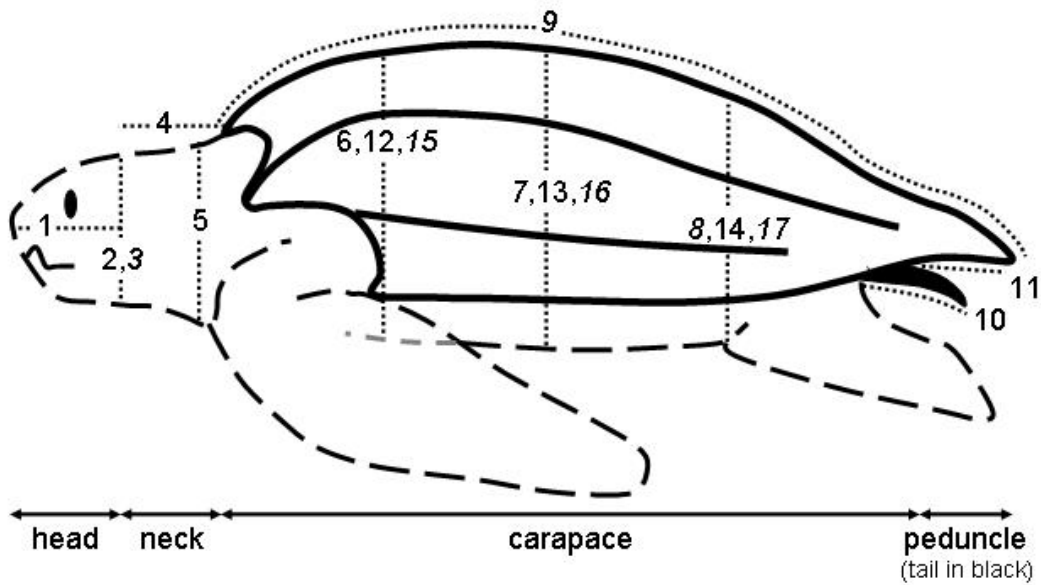

**Figure 2.** Final general linear model (after stepwise backward analysis) of body mass in gravid leatherback turtles performed on 182 individuals nesting in French Guiana between April 4 and July 27, 2005. Body Mass (kg) = -709.146 + 3.391MedianBodyCirc (cm) + 2.664SCCL (cm), $F_{2,179}$ = 1054.476, $r^2$ = 0.922, $P < 0.001$ (**model 2** in the text), where SCCL is the standard curvilinear carapace length and MedianBodyCirc is the body circumference at half of SCCL (see **Figure 1**).

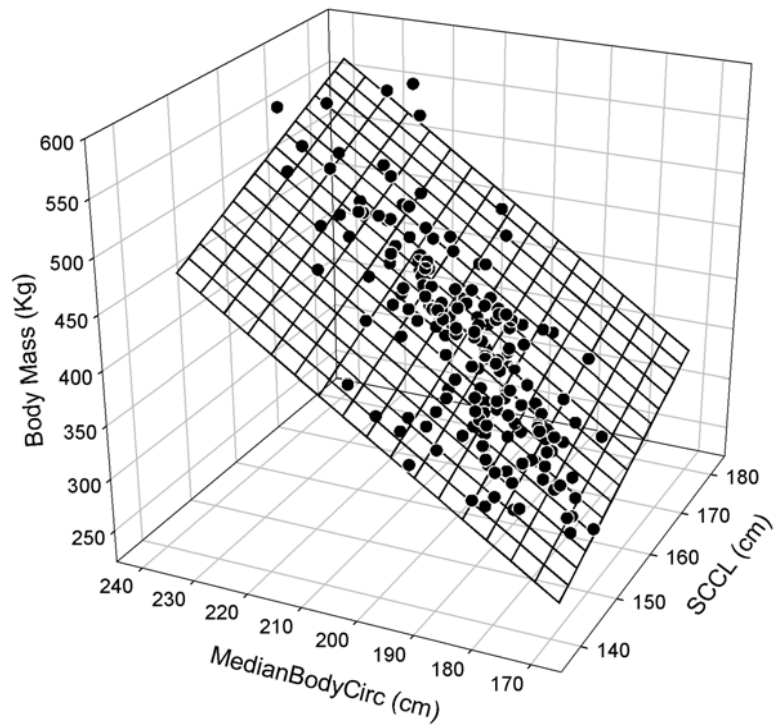

**Figure 3.** Comparison of the estimates of body mass from standard curvilinear carapace length (SCCL) and body circumference at half of SCCL (MedianBodyCirc) in gravid leatherback turtles nesting in French Guiana obtained from **model 1** (calculated from 49 individual turtles captured early in the season) and from **model 2** (calculated from 128 individual turtles captured throughout the nesting season). The line corresponds to y = x equation.

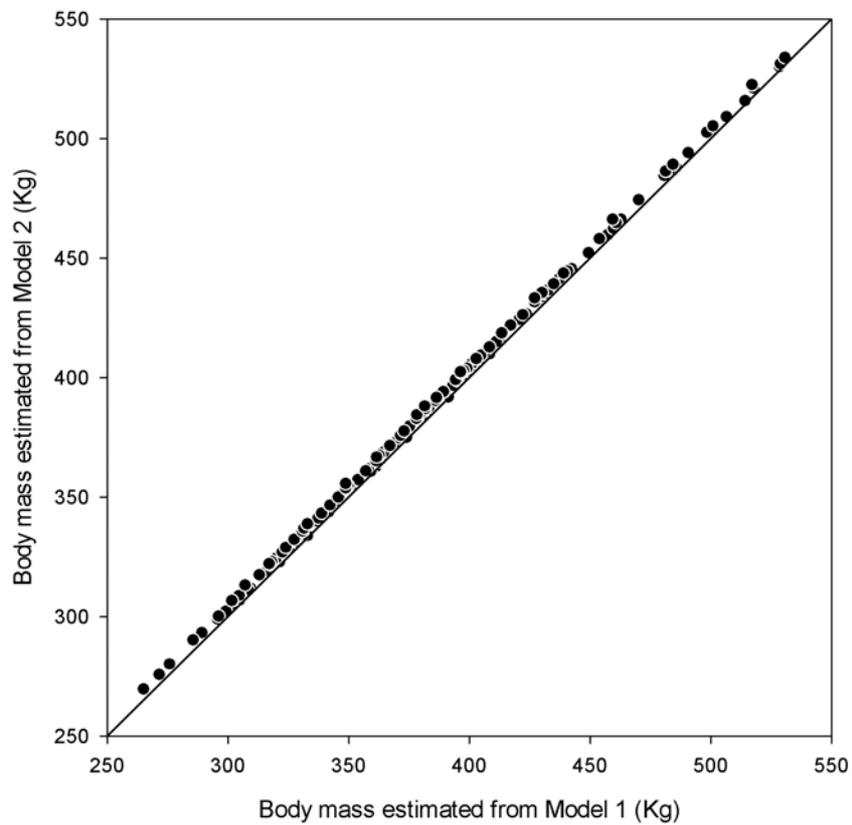

**Figure 4.** Sensibility test of the model estimating body mass from standard curvilinear carapace length (SCCL) and body circumference at half of SCCL (MedianBodyCirc) in gravid leatherback turtles nesting in French Guiana throughout the nesting season 2005. Relative changes in body mass estimated from the model (**model 2** in the text) were calculated according to changes in **(a)** SCCL and **(b)** MedianBodyCirc of 1, 2 and 5 cm.

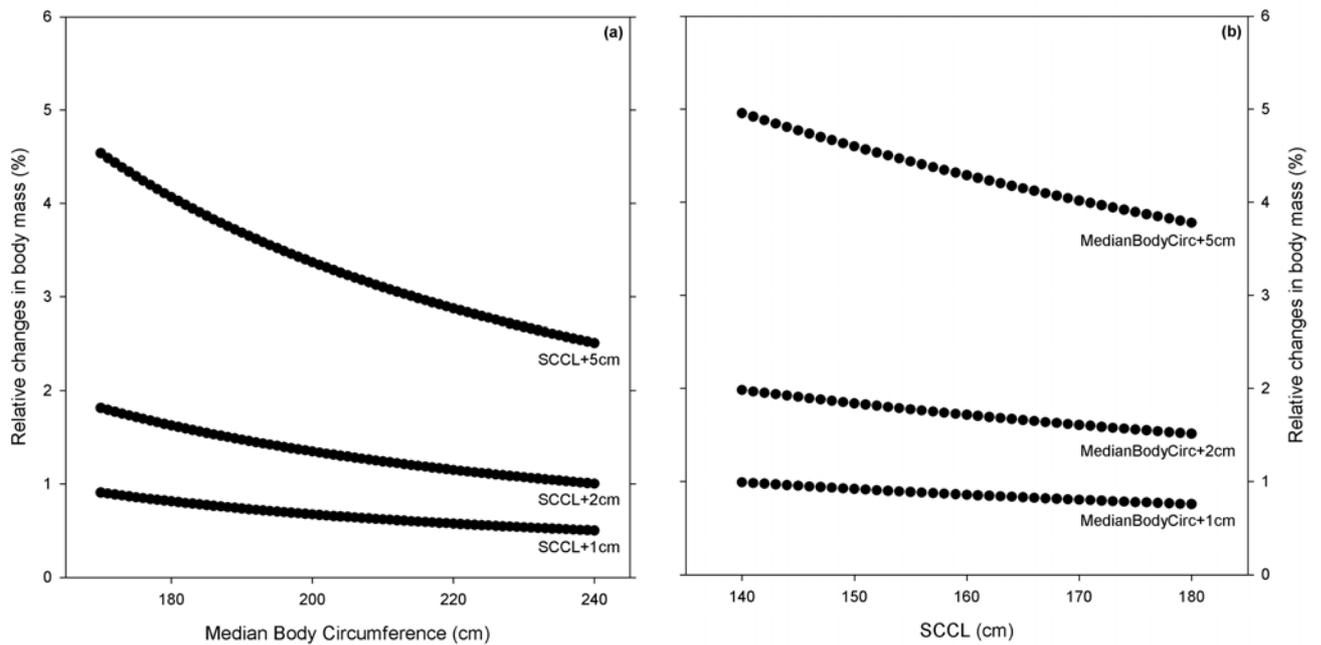

**Figure 5.** Relationship between mean body mass and mean standard curvilinear carapace length (SCCL) in leatherback turtles nesting in **(a)** Pacific Costa Rica (Eastern Pacific, n = 6 individual females estimated from Southwood et al. 1999), **(b)** St Croix Island (western Atlantic, n = 102 females estimated from Boulon et al. 1996), **(c)** Atlantic coast of Costa Rica (western Atlantic, n = 22 females estimated from Leslie et al. 1996), **(d)** and **(f)** French Guiana (western Atlantic, open circles, n = 182 females from this study and n = 15 females estimated from Girondot & Fretey 1996, respectively), and **(e)** from leatherback turtles feeding off eastern Canada (western Atlantic, n = 7 females and 6 males estimated from James et al. 2005). For **(e)** note that individuals were feeding, explaining the apparent outlier value and that no sexual dimorphism was reported, permitting to pool female and male data (James et al. 2005). The linear regression calculated for gravid leatherback on their nesting sites **(a, b, c, d, f)** was Body mass (kg) = 7.2SCCL (cm) − 769 ($F_{1,3}$ = 93.981, $r^2$ = 0.969, P = 0.002).

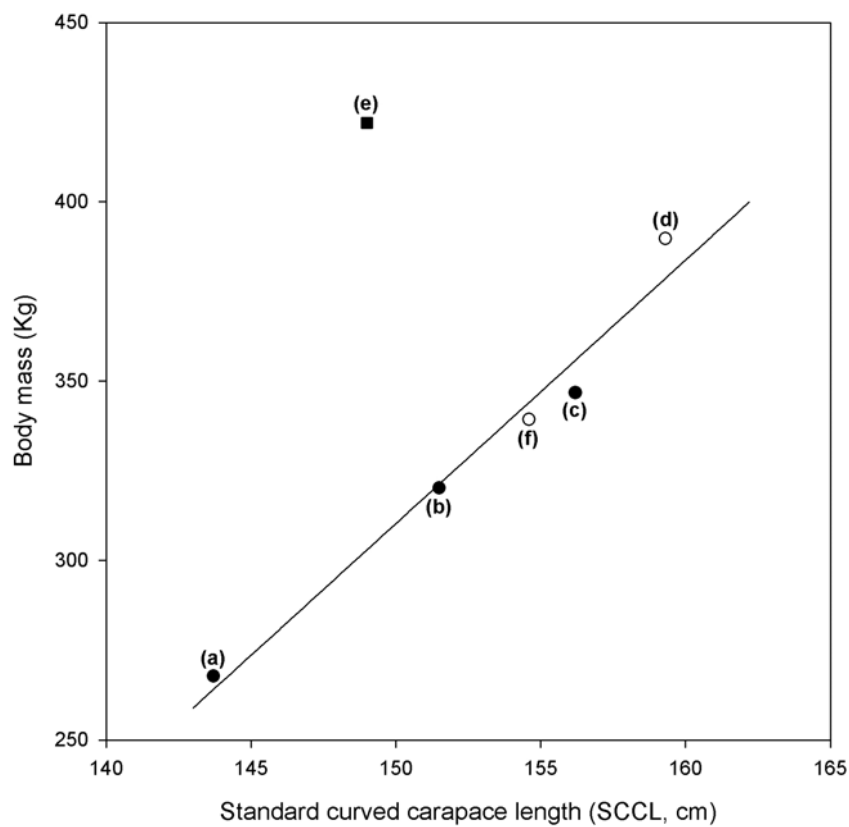